\documentclass[12pt]{book}

\usepackage[dvips]{graphicx,color}
\usepackage{makeidx,tsukuba}

\makeauthorindex
\makeindex

\begin{document}

\BookTitle{\itshape The 28th International Cosmic Ray Conference}
\CopyRight{\copyright 2003 by Universal Academy Press, Inc.}
\pagenumbering{arabic}

\newcommand{\cerenkov}{\v{C}erenkov}
\newcommand{\gr}{$\gamma$-ray}
\newcommand{\grs}{$\gamma$-rays}
\newcommand{\dg}{\ensuremath{^\circ}}
\newcommand{\chisq}{$\chi^2$}
\newcommand{\till}{$\rightarrow$}
\newcommand{\siz}{\emph{size}}
\newcommand{\wid}{\emph{width}}


\chapter{A Search for Pulsed TeV Gamma-Ray Emission from the Crab
Pulsar using the Whipple High Resolution\\ GRANITE III Camera}

\author{%
%
%
J. Kildea,$^{1,2}$ I.H.~Bond, P.J.~Boyle, S.M.~Bradbury, J.H.~Buckley,
D.~Carter-Lewis, O.~Celik, W.~Cui, M.~Daniel, M.~D'Vali,
I.de~la~Calle~Perez, C.~Duke, A.~Falcone, D.J.~Fegan, S.J.~Fegan,
J.P.~Finley, L.F.~Fortson, J.~Gaidos, S.~Gammell, K.~Gibbs,
G.H.~Gillanders, J.~Grube, J.~Hall, T.A.~Hall, D.~Hanna, A.M.~Hillas,
J.~Holder, D.~Horan, A.~Jarvis, M.~Jordan, G.E.~Kenny, M.~Kertzman,
D.~Kieda, J.~Knapp, K.~Kosack, H.~Krawczynski, F.~Krennrich,
M.J.~Lang, S.~LeBohec, E.~Linton, J.~Lloyd-Evans, A.~Milovanovic,
P.~Moriarty, D.~Muller, T.~Nagai, S.~Nolan, R.A.~Ong, R.~Pallassini,
D.~Petry, B.~Power-Mooney, J.~Quinn, M.~Quinn, K.~Ragan, P.~Rebillot,
P.T.~Reynolds, H.J.~Rose, M.~Schroedter, G.~Sembroski, S.P.~Swordy,
A.~Syson, V.V.~Vassiliev, S.P.~Wakely, G.~Walker, T.C.~Weekes,
J.~Zweerink \\
{\it
(1) McGill University, Montreal, Canada.\\
(2) The VERITAS Collaboration--see S.P.Wakely's paper} ``The VERITAS
Prototype'' {\it from these proceedings for affiliations}
}

\section*{Abstract}

\noindent We present the results of a search for pulsed TeV emission from the Crab pulsar using 97 hours of data recorded with the high-resolution GRANITE III camera of the Whipple 10~m gamma-ray telescope. 


\section{Introduction}
\noindent The Crab pulsar/nebula complex is one of the most studied objects in astrophysics with observations covering the complete detectable range of the electromagnetic spectrum, from radio waves to TeV gamma-rays. Given that pulsed emission from the pulsar is detected up to at least 10 GeV [2] and that only upper limits, well below the extrapolated spectrum, exist above $\sim$100~GeV (for example [6]), a pulsed emission cut-off necessarily lies in between. Two main emission models, the polar cap model [7] and the outer gap model [4], compete to explain high-energy gamma-ray emission from pulsars. Although both predict spectral cut-offs within the 10--100~GeV regime; the polar cap model anticipates a steep super-exponential cut-off while the outer gap model allows for a more gradual simple-exponential turnover. As such, any clear detection of pulsed gamma-ray emission by ground-based \cerenkov\ detectors would favour the outer gap model.

In 1999 the Whipple 10~m gamma-ray telescope was upgraded to hold a finely pixelated camera comprising an array of 379 PMTs with a total field-of-view of 2.6\dg. Given the improved image characterisation of the new camera, the fact that all mirrors on the 10~m reflector were re-aluminized during the upgrade, and the abundance of Crab nebula/pulsar data, it was considered appropriate to undertake a renewed search for evidence of a pulsed gamma-ray signal. 

\section{Gamma-ray Selection}
\noindent In the Crab pulsar analysis reported here a total dataset of 97.33~hours of on-source data, recorded in clear weather from January 2000 until February 2002, was examined. Prior to any gamma-ray selection analysis, all \cerenkov\ events were subjected to the standard Whipple pre-selection processing and image parameterisation using a moment-fitting analysis. Two independent analysis techniques, Supercuts [8] and Kernel analysis [1], were used to select gamma-ray images from the dataset. In order to extract the optimum number of gamma-ray events, with maximum significance above background and over the widest possible range of energies, the selection criteria for both techniques were optimised in discrete elevation and \siz\ bands using the complete dataset itself (\siz\ is a measure, in digital counts, of the total \cerenkov\ light in an image after standard image cleaning). Although generally a source of bias, optimising on the dataset itself was not considered unreasonable in this case, as TeV gamma-rays from the direction of the Crab nebula are firmly established and the present work is a search for pulsed emission. Elevation and \siz\ banding were important in order to account for the effects of both elevation and \siz\ on the gamma-ray shape parameters, and to allow for changes in collection area and peak response energy of the telescope as a function of elevation. Figure 1 presents the telescope's collection area as a function of energy at 45\dg\ and 75\dg\ elevation for the Supercuts and Kernel analyses.

The Kernel analysis was considered a useful complement to the standard Supercuts technique because its looser selection criteria allow for larger collection areas (Figure 1) at higher energies, and improved gamma-ray identification at lower energies [5], when compared to Supercuts. Using only those gamma-ray events selected by the band-specific selection criteria, described above, a temporal analysis was performed within each band and within all bands combined, to search for evidence of a periodic signal. 

\begin{figure}
  \centering
  \scalebox{0.3}
  {
    \includegraphics[draft=false,clip=false]{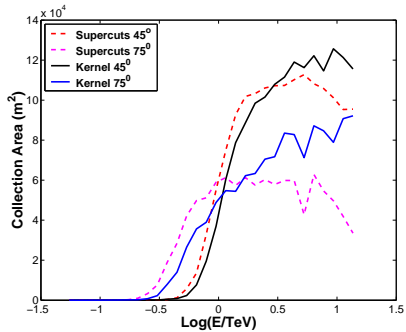}
  }
\caption{Collection area curves for the Supercuts and Kernel analyses, for the 45\dg\ and 75\dg\ elevation bands used in this analysis, with all \siz\ bands combined.}
\label{fig:elev_collect_areas}
\end{figure}

\section{Temporal Analysis}
\noindent The arrival times of \cerenkov\ events were registered by a GPS clock with an absolute resolution of 250~$\mu$s and an interpolated resolution of 0.1~$\mu$s. All times were transformed to the solar system barycentre using the JPL DE200 Planetary and Lunar Ephemeris. The phase of each event, with respect to the Crab pulsar radio ephemeris, was calculated using the Jodrell Bank monthly ephemeris updates. To test the barycentering algorithms, an optical signal of the Crab pulsar, observed on December 02, 1996, with the Multiple Mirror Telescope but passed directly through the electronics of the Whipple 10~m, was analysed. The resulting optical lightcurve showed clear evidence of a pulsed signal phase-aligned with the radio pulsations. This demonstrated the validity of the barycentering software and also the telescope's timing electronics which are essentially unchanged since 1996.


\section{Results}
\noindent No evidence for pulsed gamma-ray emission above 106~$\pm$~20~GeV was found in the gamma-ray dataset used in this work. To calculate flux upper limits, we used the method of Helene [3] to estimate 99.9\% upper limits for excess events within the pulsed phase profile seen by EGRET at lower energies [2]. That is, emission is assumed to occur in the phase range of the main pulse, phase 0.94--0.04, and the intrapulse, phase 0.32--0.43. Despite the selection of gamma-ray events by the Kernel analysis down to $\sim$106~GeV, upper limits were only determinable within the elevation and \siz\ bands for which the peak in the telescope's differential response curve was greater than 160~GeV. This was due to sky-noise problems experienced at low energies, which complicate the effective collection area determination. Studies to understand the effects of sky-noise, using simulations and an improved padding algorithm, are ongoing. The lowest energy integral upper limit in this work was estimated at $< 570 \times 10^{-13}$~cm$^{-2}$s$^{-1}$ at a peak response energy of 160~$\pm$~30~GeV (estimated using data selected in the \siz: 150--250 digital counts, elevation: 71\dg--80\dg\ band).


\vspace{-0.2in}
\section{Discussion}
\noindent To constrain the pulsed gamma-ray spectrum using the lowest energy upper limit from this work, a function of the form 
\begin{equation}
\frac{dN}{dE} = KE^{-\gamma}e^{-\frac{E}{E_0}}
\label{eqn_crab_model_spectrum}
\end{equation}
was used where $\gamma$=2.08~$\pm$~0.03 is the EGRET pulsed emission power-law spectral index [2], $E$ is the photon energy, and $E_0$ is the cutoff energy. Extrapolating the EGRET power-law to our energies using equation 1, while simultaneously constraining it with our upper limit, yields a cut-off energy of $E_0~\leq$~55~GeV for pulsed emission. Figure 2 shows our spectral point together with the model predictions and recent measurements by other groups. Although consistent with previous Whipple measurements [6], the present cut-off cannot offer discrimination between the models.

\begin{figure}
  \centering
  \scalebox{0.42}
  {
    \includegraphics[draft=false,clip=false]{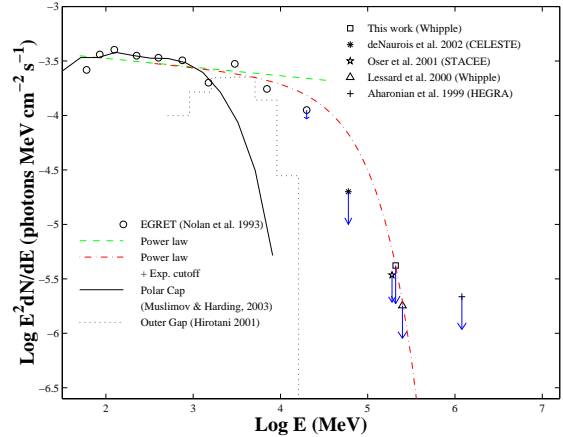}
  }
  \caption{Pulsed photon spectrum of the Crab pulsar showing the present result together with the model predictions and recent results by other groups.}
  \label{fig:pulsar_cuts_area}
\end{figure}

\section{Acknowledgements}
We acknowledge the technical assistance of E. Roache and J. Melnick. J. Kildea is grateful for useful discussions with A.K. Harding, K. Hirotani, and D.A. Smith. This research is supported by grants from the U.S. Department of Energy, by Enterprise Ireland and by PPARC in the UK.

\section{References}

\noindent 1. Dunlea S. et al. 2001, ICRC Hamburg, vol 7, 2939\\
2. Fierro J.M. 1995, PhD thesis, Stanford University\\
3. Helene O. 1983, Nuclear Instruments and Methods, 212, 319\\
4. Hirotani K. and Shibata S. 2001, MNRAS, 325, 1228\\
5. Kildea J. 2002, PhD thesis, National University of Ireland\\
6. Lessard R.W. et al. 2000, ApJ, 531, 942\\
7. Muslimov A.G and Harding A.K. 2003, ApJ, 588, 430\\
8. Punch M. 1991, ICRC Dublin, vol 1, 464\\

\endofpaper
\end{document}